\documentclass[reprint,superscriptaddress,
 amsmath,amssymb,
 aps,prb,longbibliography,floatfix]{revtex4-1}
\usepackage{amsmath,amssymb}
\usepackage{graphicx}
\usepackage{color}
\usepackage{bm}
\usepackage{hyperref}
\hypersetup{
colorlinks=true,
urlcolor= blue,
citecolor=blue,
linkcolor= blue,
bookmarks=true,
}

\newcommand*\diff{\mathop{}\!\mathrm{d}}
\renewcommand{\vec}[1]{\bm{#1}}

\begin{document}

\title{Anomalous impurity-induced charge modulations in black phosphorus}

\affiliation{Department of Physics, Yonsei University, Seoul 03722, Republic of Korea}
\affiliation{Max Planck Institute for Chemical Physics of Solids, 01187 Dresden, Germany}
\affiliation{Institute for Theoretical Physics III, University of Stuttgart, 70550 Stuttgart, Germany}
\affiliation{Department of Physics and Center for Nano Materials, Sogang University, Seoul 04107, Republic of Korea}

\author{Byeongin Lee}
\thanks{These authors contributed equally to this work.}
\affiliation{Department of Physics, Yonsei University, Seoul 03722, Republic of Korea}

\author{Junho Bang}
\thanks{These authors contributed equally to this work.}
\affiliation{Department of Physics, Yonsei University, Seoul 03722, Republic of Korea}

\author{Sayan Banerjee}
\affiliation{Institute for Theoretical Physics III, University of Stuttgart, 70550 Stuttgart, Germany}

\author{Jo$\mathrm{\tilde{a}}$o Augusto Sobral}
\affiliation{Institute for Theoretical Physics III, University of Stuttgart, 70550 Stuttgart, Germany}

\author{Young Woo Choi}
\affiliation{Department of Physics and Center for Nano Materials, Sogang University, Seoul 04107, Republic of Korea}

\author{Claudia Felser}
\affiliation{Max Planck Institute for Chemical Physics of Solids, 01187 Dresden, Germany}

\author{Mathias S. Scheurer}
\affiliation{Institute for Theoretical Physics III, University of Stuttgart, 70550 Stuttgart, Germany}

\author{Jian-Feng Ge}
\email[Corresponding author:~]{Jianfeng.Ge@cpfs.mpg.de}
\affiliation{Max Planck Institute for Chemical Physics of Solids, 01187 Dresden, Germany}

\author{Doohee Cho}
\email[Corresponding author:~]{dooheecho@yonsei.ac.kr}
\affiliation{Department of Physics, Yonsei University, Seoul 03722, Republic of Korea}

\date{\today}

\begin{abstract}

We observe anomalous charge modulations induced by ionized indium impurities on the surface of the semiconductor black phosphorus by scanning tunneling microscopy (STM). When the impurities are switched into a negatively charged state by the STM tip, periodic charge modulations emerge around the impurity center, but strictly confined by the nanoscale impurity potential. These modulations form a distorted triangular pattern, whose periodicity remains unchanged in a wide range of positive bias. Furthermore, these local charge orders exhibit an anisotropy opposite to that expected based on the anisotropy of the Fermi surface, challenging a simple band-structure interpretation. Our experiment demonstrates the possibility of creating and manipulating macroscopic charge orders through impurity engineering.
\end{abstract}

\maketitle 

\section{\label{sec:intro} I. Introduction}

Impurities in low-dimensional electronic materials often act as localized perturbations that reorganize the surrounding charge density or atomic lattice. Such rearrangements can appear as standing-wave-like charge density modulations \cite{Hasegawa1993-zj,Crommie1993-ux,Hoffman2002-xz}, localized impurity states \cite{kitchen2006atom}, or impurity-induced lattice distortions \cite{reticcioli2017polaron, zhang2020nanoscale}, depending on the strength and spatial extent of the impurity potential. Understanding these responses is essential for determining how impurities enable tuning material properties and for revealing the mechanisms behind impurity-driven electronic behavior \cite{balatsky2006impurity, Evers2008-cw, yazyev2007defect, Park2024Oct}. Despite extensive studies on impurity-induced electronic signatures in various materials, there are still open questions concerning the microscopic mechanism by which an individual impurity reshapes the nearby electronic environment, particularly in systems with strong structural or electronic anisotropy.

Black phosphorus (BP) provides an advantageous platform for investigating these effects. Its puckered structure generates markedly anisotropic effective masses, yielding distinct Fermi velocities \cite{Kim2015-xi,Baik2015-fi} and screening behaviors depending on crystallographic axes \cite{kiraly2017probing, Kiraly2019-cg}. With scanning tunneling microscopy (STM), one can directly tune the ionization state of a surface impurity and thereby control the associated Coulomb (or scattering) potential \cite{Teichmann2008-fn}. This enables real-space visualization of how electrons respond to an impurity whose charge state and potential profile can be manipulated with an atomic precision. Previous studies have examined impurity charging and scattering in BP \cite{fang2022electronic, zouFriedel}, but it remains uncertain whether an ionized impurity can induce anisotropic charge modulations expected from the intrinsic band anisotropy.

Here, we employ STM to control the charge state of individual indium clusters on BP and uncover anomalous charge modulations. These modulations form a distorted triangular order, remain sharply confined within the tip-induced band-bending region, and exhibit an energy-independent wavevector. Notably, their spatial anisotropy follows the zigzag direction, contrary to expectations based on the effective-mass anisotropy and known screening behavior of BP \cite{kiraly2019anisotropic}. These findings show that the tunability of impurity charge states provides a powerful means to elucidate impurity-induced electronic modulations unaccounted for by standard scattering or screening models.

\section{II. Results and Discussion}
%Figure 1
\begin{figure*}[t]
\includegraphics{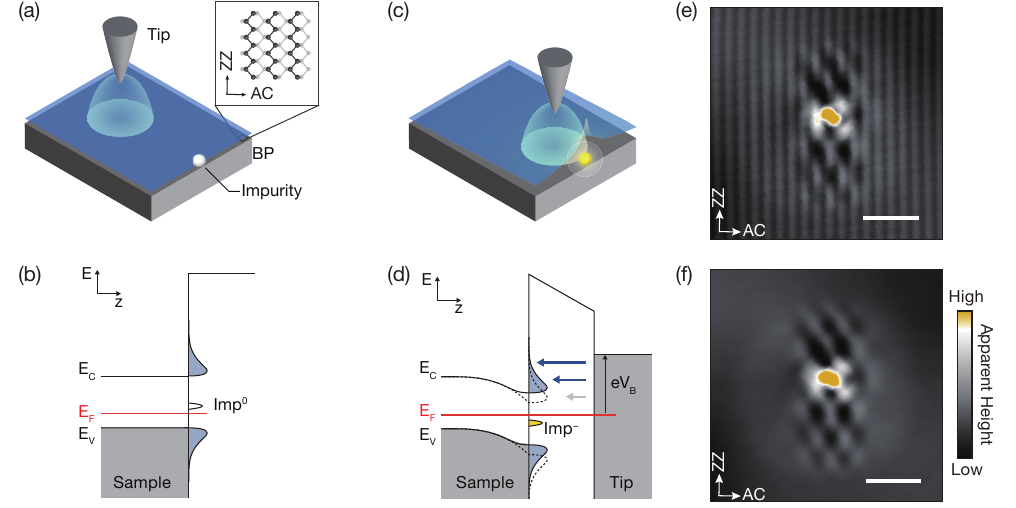}
\caption{Charge modulations near ionized adatom impurities. (a) Schematic illustration of the experimental setup. The STM tip is brought close to the sample (gray), and the electric field between them (cyan) locally penetrates the sample. A neutral impurity (white ball) is placed on the surface. When the tip is away from the impurity, the surface electrons (blue) remains spatially uniform. The inset shows the crystal structure of the (001) surface of the black phosphorus (BP) samples used. The darker and brighter balls represent surface and subsurface phosphorus atoms, respectively. (b) Local energy diagram near the impurity for the case in (a). $E_\mathrm{C}$ and $E_\mathrm{V}$ denote band edges of the conduction (white) and valence (gray) bands, respectively. The red line indicates the Fermi level, $E_\mathrm{F}$. The blue shades illustrate the density of states on the surface. The impurity level (white) above $E_\mathrm{F}$ indicates the impurity’s charge-neutral (Imp$^0$) state. (c) Ionization (yellow ball) occurs when the local electric field reaches the impurity, and the impurity’s Coulomb potential disturbs locally the energy of surface electrons (blue). (d) Local energy diagram near the impurity for the case in (c). $V_\mathrm{B}$ is the applied bias voltage. Ionization (Imp$^-$) is represented by the impurity level (yellow) moving below the $E_\mathrm{F}$. While the dashed curves illustrate tip-induced band bending, the Coulomb potential of ionized impurity further lifts the bands (solid curves). The blue arrows represent the tunneling channels, with their lengths indicating the tunneling probabilities. (e) STM image taken near indium adatoms showing charge modulations. Setup conditions: $V_\mathrm{B}$ = 0.7 V, $I_\mathrm{set}$ = 0.1 nA. (f) Fourier-filtered image of (e), highlighting the charge modulations. Scale bar, 2 nm.}
\label{fig:1}
\end{figure*}

We begin with a two-dimensional electron system on the surface of black phosphorus. BP is a predominantly p-type semiconductor, and its low-energy electronic band structure can be approximately modeled as simple parabolic dispersions for both conduction and valence bands \cite{Kim2015-xi,Baik2015-fi}. The (001) surface of BP exhibits a puckered honeycomb lattice structure [inset of Fig.~\ref{fig:1}(a)], resulting in two distinct directions, designated as zigzag (ZZ) and armchair (AC). One peculiar characteristic of BP’s electronic structure is its anisotropy along these two directions, manifested by a substantial difference in the effective mass of the conduction band electrons $m_\mathrm{AC}^* = 0.07 m_e$ for the armchair direction and $m_\mathrm{ZZ}^* = 1.0 m_e$ for the zigzag direction \cite{Chen2019-el}, where $m_e$ is the free electron mass. Next, we introduce impurities on BP by depositing indium atoms, which form either clusters on the surface with a diameter of 1 $\sim$ 5 nm, or triangular islands with size over 5 nm \cite{SI}. 

We then proceed to switch the ionization state of the surface impurities. A surface impurity creates a strong Coulomb potential, but only when the impurity is charged. A neutral impurity, otherwise, would only weakly perturb the electron density around it. Experimentally, it is well established that the ionization state of an impurity in a semiconductor can be controlled by STM \cite{Repp2004-wy, Teichmann2008-fn, Marczinowski2008-qh}, as illustrated in Fig.~\ref{fig:1}. When an STM tip approaches the sample surface, the strong electric field (typically on the order of 1 V/nm) exerted by the tip extends toward the sample. If screening is insufficient, this field penetrates the local surface area under the tip [Fig.~\ref{fig:1}(a)], modifying the local potential inside the sample. This often results in a rigid shift of the conduction and valence band edges [dashed lines in Fig.~\ref{fig:1}(d)], known as tip-induced band bending (TIBB) \cite{Feenstra1987-du}. When the impurity is far from the tip, it stays neutralized with the surrounding background [Fig.~\ref{fig:1}(b)]. However, when the spatial extension of the tip’s field reaches the impurity [Fig.~\ref{fig:1}(c)] and shifts down the lowest unoccupied impurity level across $E_\mathrm{F}$, the impurity becomes negatively ionized. This ionization subsequently enhances the impurity’s Coulomb potential, which in turn lifts the bands at the surface. The combination of tip- and impurity-induced band bending effects have been understood in previous reports \cite{Wijnheijmer2009-fn, Song2012-of}, as shown by comparing the solid and dashed curves in Fig.~\ref{fig:1}(d). In this case, as the tunneling channels with higher transmission probability (blue arrows) align closer to a peak in the surface density of states (blue shades) \cite{Golias2016-iq}, the tunneling current is instantaneously increased compared to the case before ionization \cite{SI}. In constant-current mode, the feedback control of STM immediately retracts the tip. It is then anticipated that this change in current occurs roughly at the same distance from the impurity in all directions, manifesting as a disk-like feature with enhanced contrast in the STM image \cite{Repp2004-wy, Teichmann2008-fn, Marczinowski2008-qh}. As illustrated in Fig.~\ref{fig:1}(e), this is indeed what we observed around a dimer impurity. Behind the vertical stripes originating from the zigzag ridges and valleys, we observe a faint disk-like feature around the impurity. The area inside the disk exhibits a larger apparent height at positive bias voltages, which is better discernible after filtering out the surface corrugation in Fig.~\ref{fig:1}(f) \cite{SI}. 

\subsection{A. Charge modulations near ionized impurities}

The primary finding of this study is the discovery of additional modulations in the vicinity of the impurity, which exist within the commonly observed disk-like feature. These modulations can be clearly visualized as oscillations in both filtered and unfiltered images in Figs.~\ref{fig:1}(e) and \ref{fig:1}(f), as well as in differential conductance images taken simultaneously \cite{SI}. They appear to form an approximately triangular lattice that is distorted because of the underlying orthorhombic surface Bravais lattice, leading to a centered rectangular charge modulation; additional distortions are visible at the disk edge and near the central impurity. One important observation is that these modulations are strictly confined to the disk. To demonstrate this, we change the disk radius by varying the bias voltage or the tip-sample distance, as TIBB depends on the electric field between the tip and sample.

%Figure 2
\begin{figure}[t]
\includegraphics{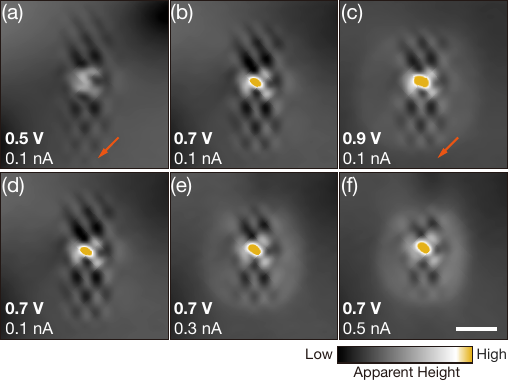}
\caption{Charge modulations controlled by the electric field between tip and sample. (a)--(c) Fourier-filtered STM images taken at a fixed current setpoint but with varying bias voltages. As the bias voltage increases, the size of the disk decreases. The orange arrows indicate that a charge modulation peak shown in (a) disappears at the same location in (c). (d)--(f) Fourier-filtered STM images taken at a fixed bias but with varying current setpoints. Scale bar, 2 nm.}
\label{fig:2}
\end{figure}

As shown in Fig.~\ref{fig:2}, the disk shrinks in size with an increasing bias or a decreasing tip-sample distance (increasing current setpoint). This band-bending behavior confirms that the impurity is negatively charged when the tip is positioned inside the disk \cite{SI}. As the disk size decreases, the area that contains the modulations also shrinks accordingly. This confinement is clearly visible when comparing the same field of view, for example, at a fixed current setpoint but increasing bias in Figs.~\ref{fig:2}(a)--(c). In Fig.~\ref{fig:2}(a), the modulations extend to over 4 nm away from the impurity along the zigzag direction; a density peak appears at the location indicated by the arrow. In contrast, in Fig.~\ref{fig:2}(c) the modulations disappear at the same location as the disk shrinks. We observe the same phenomenon when comparing images taken at a fixed bias of 0.7 V but with increasing current setpoint [Figs.~\ref{fig:2}(d)--(f)]. These observations rule out the possibility of structural deformation and suggest an electronic origin of the modulations. 

\subsection{B. Possible origin of the charge modulations}
To discuss possible electronic mechanisms, we first consider Friedel oscillations, often referred to as quasiparticle interference (QPI) and widely utilized in determining electronic band structures by STM \cite{Hasegawa1993-zj,Crommie1993-ux,Hoffman2002-xz}. Specifically, the oscillations should have a scattering wavevector $\vec{q}(E)=\vec{k}_\mathrm{f}(E) - \vec{k}_\mathrm{i}(E)$, connecting the wavevectors  $\vec{k}_\mathrm{i}(E)$  and $\vec{k}_\mathrm{f}(E)$ of the initial and final states at a certain energy $E$ in the energy bands, respectively. A Fourier transform of the real-space oscillation pattern at each energy reveals the dispersion relation $\vec{q}(E)$, which can be used to reconstruct the energy bands. In our case, an anisotropic parabolic dispersion mode is expected from the first-principle calculations and photoemission experiments \cite{Kim2015-xi, Chen2019-el}.

%Figure 3
\begin{figure}[tb]
\includegraphics{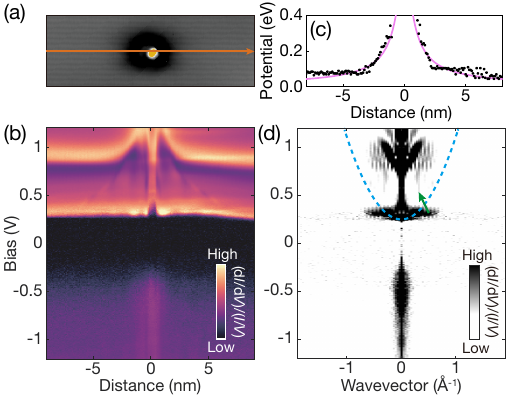}
\caption{Energy dependence of charge-modulation periodicity. (a) STM image taken near a two-adatom impurity. Setup conditions: $V_\mathrm{B}$ = 1.2 V, $I_\mathrm{set}$ = 0.1 nA. (b) The normalized differential conductance $(\diff{I}/\diff{V})/(I/V)$ spectra along the line marked by the arrow in (a). Setup conditions: $V_\mathrm{B}$ = 1.2 V, $I_\mathrm{set}$ = 0.2 nA. (c) Band edge (black dots) appearing near $V_\mathrm{B}$ = 0.8 V in (b), extracted by finding the bias that corresponds to a maximum in the first derivative of each spectrum above 0.7 V. The impurity potential (purple) is estimated by a screened Coulomb potential \cite{SI}. (d) Line-wise Fourier transform of (b). The periodic feature of the charge order is marked with the green arrow. The blue dashed curve corresponds to the expected dispersion of quasiparticle interference along the zigzag direction in the conduction band.}
\label{fig:3}
\end{figure}

To obtain the wavevector $\vec{q}(E)$ associated with our observed modulations, we measure the differential conductance along a line in the zigzag directions where modulations appear [Fig.~\ref{fig:3}(b)] and then perform a line-wise Fourier transform [Fig.~\ref{fig:3}(d)]. We first note the conduction band that starts at 0.3 V, and slightly bends upwards right on top of the impurity; this results in a flat dark feature in the Fourier transform in Fig.~\ref{fig:3}(d). 
We then focus on the bias range of 0.5 V to 1.2 V where the modulations appear. The spatial extent of the modulations diminishes due to the reduced disk radius with increasing bias [Fig.~\ref{fig:3}(b)], as also indicated by the V-shaped dark feature with multiple sidebands in Fig.~\ref{fig:3}(d) due to quantum confinement effects. Most importantly, we note an almost constant wavevector $\lvert\vec{q}\rvert \sim$~0.32~\r{A}$^{-1}$ that corresponds to the observed modulations. 
This energy-independent mode clearly deviates from the expected parabolic dispersion [blue dashed line in Figs.~\ref{fig:3}(d)], thus indicating that trivial Friedel oscillations would not suffice to explain the observed modulations. Interestingly, it is known that the QPI signal can be significantly influenced by quantum geometric form factors~\cite{Zhang2019Nov}. In this case, we do find the possibility that an apparent energy-independent mode appears in our simulated QPI. Specifically, non-local impurity potentials with selective orbital couplings suppress scatterings along the zigzag direction \cite{SI}.

Furthermore, our results show a spatial anisotropy opposite to what one would expect from the Fermi surface. Due to the anisotropy of the effective masses along the zigzag and armchair directions, the Fermi surface (of doped BP) has an elliptical shape elongated along the zigzag directions \cite{Kim2015-xi, Ehlen2018-zq}. This means that the screening of the extra charges by band electrons should have a longer decay length along the armchair direction \cite{Kiraly2019-cg}, and that any charge modulation, e.g., originating from Fermi surface nesting, should have a longer length scale along the armchair direction. Surprisingly, upon close inspection of the charge modulations in Fig.~\ref{fig:1}(f), we find that the individual density peaks have an elliptical shape elongated along the zigzag direction instead of the armchair direction. In addition, the modulations extend to the boundary of the disk along the zigzag direction but are more confined spatially along the armchair direction. Inside the disk, we also observe a smooth transition from modulations to a relatively uniform background when moving toward the disk edge along the armchair direction. Both the spatial extent of the modulations and the shape of the individual density peaks are intrinsic to the charge orders and irrespective of the appearance of the impurity, as confirmed by imaging charge modulations around different impurities and nanoislands \cite{SI}. While local impurities can mix bands in a way consistent with a shorter wavelength along the armchair direction~\cite{Baik2015-fi, Jang2019-jn}, there is no indication of such a strong redistribution across the band gap, which stays open with a nearly constant size when approaching the impurity [Fig.~\ref{fig:3}(b)]. 

We also consider whether enhanced interaction effects give rise to the charge modulations. The presence of a strong, localized impurity potential can trap electrons and one might hypothesize that it simultaneously reduces the screening of Coulomb interactions between them, potentially stabilizing a charge-ordered phase within the confined region. This interpretation is supported by our observation that the modulations being strictly confined by the potential of the ionized impurity (Fig.~\ref{fig:2}). From the differential conductance plot [Fig.~\ref{fig:3}(b)], we can extract the impurity potential, which follows the expected screened Coulomb form at the dielectric-vacuum interface~\cite{Teichmann2008-fn, Ebert1996-zh}. Assuming a tip-sample distance of 0.5 nm, and that each atom in the indium dimer acquires one electron charge after ionization \cite{SI}, we obtain a potential profile shown as the curve in Fig.~\ref{fig:3}(c). The expected impurity potential aligns well with a band edge appearing at a bias of $\sim$~0.8 V, indicating the effect of the impurity potential acting on this band.

Finally, while the charge modulations exhibit features reminiscent of a Wigner crystal~\cite{Tsui2024-iq}—including a distorted triangular lattice incommensurate with the host lattice—quantitative analysis reveals fundamental discrepancies: if each peak in Fig.~\ref{fig:1}(f) in the disk corresponded to a ‘frozen’ electron, we would have an electron density of 5×10$^{13}$ cm$^{-2}$; this corresponds to a Wigner-Seitz radius $r_\mathrm{s} \approx 0.5$, which is at least an order of magnitude lower than what is required for crystallization \cite{Tanatar1989-wh, Chui1995-bj, Egger1999-fy}. The unique appearance of these charge modulations near indium impurities—absent for phosphorus vacancies \cite{kiraly2017probing}, ionized potassium \cite{Kiraly2019-cg}, or cobalt adatoms \cite{Kiraly2018-va}—suggests that the underlying mechanism likely involves an intricate interplay between (at least a subset of) the non-local impurity potential, quantum geometric effects, and enhanced local electronic correlations.

\subsection{C. Manipulation of charge modulations}
\begin{figure}[tb]
\includegraphics{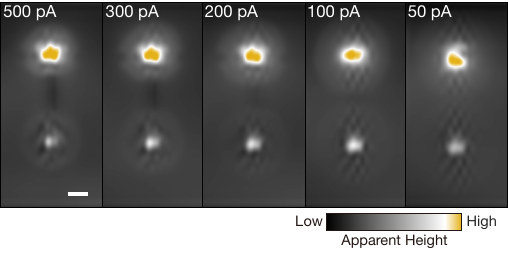}
\caption{Fourier-filtered STM images taken around two nearby cluster impurities. The bias is fixed at 0.6 V, while the current setpoint decreases from 500 pA to 50 pA. The charge modulations around both impurities are initially separated at 500 pA, and gradually expand their extensions and start to merge at 50 pA. Scale bar, 2 nm.}
\label{fig:4}
\end{figure}

To complete the story, we demonstrate how the impurity potential provides a microscopic means of manipulating the charge orders. To do so, we monitor the evolution of charge modulations around two nearby clusters (Fig.~\ref{fig:4}) while controlling the spatial profile of the impurity potentials.
%Figure 4
As previously demonstrated in Fig.~\ref{fig:2}, we can expand the spatial extent of the impurity potential by reducing the size of the TIBB disk. With a decreasing current setpoint, we find that the charge modulations around the two clusters extend toward each other. For example, at a setpoint of 50 pA, the modulations eventually merge. This microscopic enlargement of the charge modulations implies that, through impurity engineering such as nano-patterning \cite{Han2014-dg}, a macroscopic phase of such charge order can form with an appearance analogues to a charge density wave.

\section{III. Conclusions}
In conclusion, we present a direct observation of anomalous charge modulations near indium impurities in black phosphorus. The charge modulations appear as a distorted triangular lattice, with an apparent anisotropy opposite to that expected based on the Fermi surface’s anisotropy. This unexpected spatial anisotropy and the energy-independent behavior of these modulations cannot be explained by standard scattering between the quasi-particle energy bands. Instead, it might be the result of non-trivial wave-function effects, the interplay of localization and strong-coupling physics, or a combination of both.
Our work establishes a versatile experimental platform for exploring how impurities can stabilize intriguing charge orders in a system with strong intrinsic anisotropy.

\begin{acknowledgments}
\section{Acknowledgments}
We thank S. Wirth and J. van Ruitenbeek for valuable discussions.
B. L., J. B., and D. C. were supported by the National Research Foundation of Korea (NRF) grant funded by the Korea government (No. RS-2023-00251265, RS-2024-00337267, and RS-2024-00442483). and the Industry-Academy joint research program between Samsung Electronics and Yonsei University. Y. W. C. was supported by the NRF grant funded by the Korea government (No. RS-2024-00441954). S.B.~and M.S.~acknowledge funding by the European Union (ERC-2021-STG, Project 101040651---SuperCorr). Views and opinions expressed are however those of the authors only and do not necessarily reflect those of the European Union or the European Research Council Executive Agency. Neither the European Union nor the granting authority can be held responsible for them.
\end{acknowledgments}

\bibliographystyle{apsrev4-2-title}
\bibliography{ref}

\end{document}

\title{\textmd{Supplementary Material for} \protect\\[10pt] Anomalous impurity-induced charge modulations in black phosphorus}
% set order of affiliations first
\affiliation{Department of Physics, Yonsei University, Seoul 03722, Republic of Korea}
\affiliation{Max Planck Institute for Chemical Physics of Solids, 01187 Dresden, Germany}
\affiliation{Institute for Theoretical Physics III, University of Stuttgart, 70550 Stuttgart, Germany}
\affiliation{Department of Physics and Center for Nano Materials, Sogang University, Seoul 04107, Republic of Korea}
% set order of authors
\author{Byeongin Lee}
\thanks{These authors contributed equally to this work.}
\affiliation{Department of Physics, Yonsei University, Seoul 03722, Republic of Korea}

\author{Junho Bang}
\thanks{These authors contributed equally to this work.}
\affiliation{Department of Physics, Yonsei University, Seoul 03722, Republic of Korea}

\author{Sayan Banerjee}
\affiliation{Institute for Theoretical Physics III, University of Stuttgart, 70550 Stuttgart, Germany}

\author{Jo$\mathrm{\tilde{a}}$o Augusto Sobral}
\affiliation{Institute for Theoretical Physics III, University of Stuttgart, 70550 Stuttgart, Germany}

\author{Young Woo Choi}
\affiliation{Department of Physics and Center for Nano Materials, Sogang University, Seoul 04107, Republic of Korea}

\author{Claudia Felser}
\affiliation{Max Planck Institute for Chemical Physics of Solids, 01187 Dresden, Germany}

\author{Mathias S. Scheurer}
\affiliation{Institute for Theoretical Physics III, University of Stuttgart, 70550 Stuttgart, Germany}

\author{Jian-Feng Ge}
\email[Corresponding author:~]{Jianfeng.Ge@cpfs.mpg.de}
\affiliation{Max Planck Institute for Chemical Physics of Solids, 01187 Dresden, Germany}

\author{Doohee Cho}
\email[Corresponding author:~]{dooheecho@yonsei.ac.kr}
\affiliation{Department of Physics, Yonsei University, Seoul 03722, Republic of Korea}

\maketitle 
\section{Methods: Sample preparation and STM measurement}
Commercial single crystals of black phosphorus (HQ-graphene) were used for STM measurement. The bulk crystals are cleaved at room temperature in the ultra-high vacuum with a base pressure of 1×$10^{-10}$ Torr. Subsequently, indium is deposited onto the cleaved surface at room temperature using a Knudsen cell. Without post-annealing, the samples were transferred to the precooled STM head. We performed STM measurements using a commercial low-temperature STM (USM1200LL, Unisoku Co., Ltd) at 4.2 K. We used mechanically sharpened PtIr tips, whose condition was verified on the Au(111) surface. STM images were acquired in constant current mode, while differential conductance spectra were obtained using a standard lock-in technique with a modulation frequency of 913 Hz and a peak-to-peak amplitude of 10 mV.

%Figure S1
\begin{figure}[b]
\includegraphics{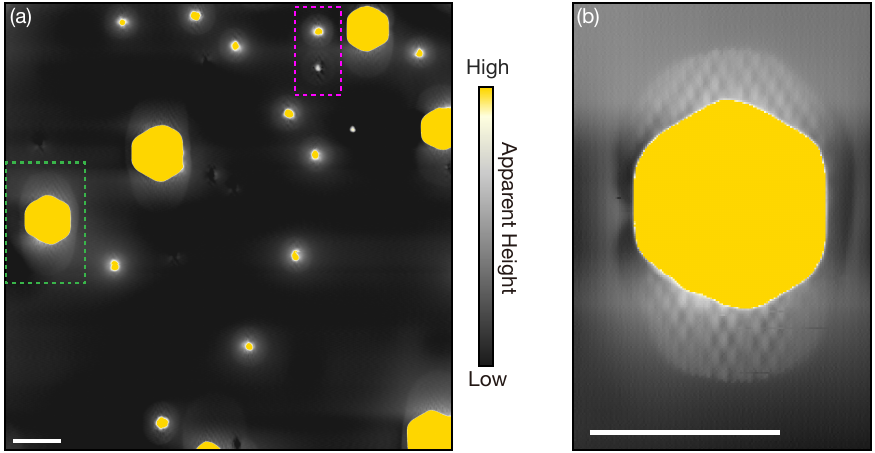}
\caption{Charge modulations around surface impurities of various sizes on BP. (a) STM image of various-sized surface impurities such as adatoms, nanoclusters, and nanoislands in a large field of view. All nanoislands have identical heights regardless of their lateral sizes due to the energy gain from the quantum confinement effect along the perpendicular direction to the surface. The magenta dashed rectangle indicates where the images in Fig. 4 were taken. (b) Zoom-in STM image of the island marked by the green dashed rectangle in (a). Charge modulations similar to those in the main text can be observed. Scale bar, 10 nm. Setup conditions: (a) $V_\mathrm{B}$ = 1.0 V, $I_\mathrm{set}$ = 0.1 nA; (b) $V_\mathrm{B}$ = 0.6 V, $I_\mathrm{set}$ = 0.1 nA.}
\label{fig:S1}
\end{figure}

\section{Supplementary Note 1: tip-induced band bending}
\begin{figure}[tb]
\includegraphics{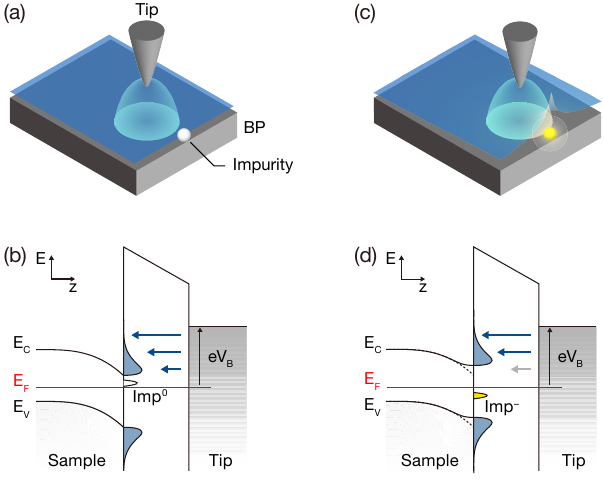}
\caption{Comparison of the tunneling current before and after ionization. Schematic illustrations of the situations before (a) and after (b) ionization. The impurity is neutral (white ball) but subsequently ionized (yellow) after a tip movement during the scan. The impurity’s Coulomb potential disturbs locally the energy of surface electrons (blue) after ionization. (c)and (d) Local energy diagram near the impurity for the case in (a) and (b), respectively. Ionization is represented by a shift of the impurity level across  $E_\mathrm{F}$. This shift is usually 10 $\sim$ 100 meV but exaggerated here in the energy (vertical) axis for illustration. The dashed curves in (d) are identical to the solid ones in (c) due to tip-induced band bending. The solid curves in (d) are the bands lifted by the impurity’s Coulomb potential based on the dashed curves. The arrows represent the tunneling probability. The blue arrows indicate allowed tunneling channels, while the gray arrow indicates a forbidden tunneling channel after band lifting.}
\label{fig:S2}
\end{figure}

The phenomenon of impurity (de)ionization in a scanning tunneling microscopy (STM) experiment is commonly described by tip-induced band bending (TIBB). This local band bending potential alters the discrete impurity levels, causing the lowest unoccupied (highest occupied) level to cross the Fermi level ($E_\mathrm{F}$), thereby gaining (losing) charge at the defect. In this section, we explain: (1) the ionization process leads to an immediate increase in tunneling current in our case, and (2) the dependences of disk diameter on bias voltage and tip-sample distance.
First, we elucidate how impurity ionization enhances the tunneling current. Using the Wentzel–Kramers–Brillouin approximation, the tunneling current of an STM junction is a transmission-weighted integration of the density of states of the tip $\rho_\mathrm{t}$ and sample $\rho_\mathrm{s}$ \cite{Hellenthal2013-cl}

\begin{equation}
I = \int ^{eV_\mathrm{B}} _0 \rho_\mathrm{s}(E)\rho_\mathrm{t}(E-eV_\mathrm{B})T(E)\diff{E},
\tag{S1}\label{eq:S1}
\end{equation}
where $e$ is the charge of an electron, and the transmission probability $T(E)$ decays exponentially with energy \cite{Hamers2000-lp},
\begin{equation}
T(E) \propto \exp\Bigg (-\sqrt{\frac{W_\mathrm{t}+W_\mathrm{s}+eV_\mathrm{B}}{2}-E}\Bigg ).
\tag{S2}\label{eq:S2}
\end{equation}
Here $W_\mathrm{t}$ and $W_\mathrm{s}$ are the work function of the tip and sample, respectively. The transmission probability $T(E)$ is depicted by the arrows in Fig. 1 and Fig.~\ref{fig:S2}; their lengths schematically indicate the decay of $T(E)$ away from the tip’s $E_\mathrm{F}$.
%\clearpage
%Figure S2
Assuming a constant $\rho_\mathrm{t}$, if $\rho_\mathrm{s}$ has a peak around a certain energy $E_0$, resonant tunneling is established when the peak of $T(E)$ aligns with the peak in $\rho_\mathrm{s}$, leading to a maximum current. This resonant tunneling, usually via impurity levels to the bulk Femi level $E_\mathrm{F}$, is a common interpretation of a sudden current change \cite{Teichmann2008-fn}. This interpretation is not likely applicable to our case where the bulk $E_\mathrm{F}$ lies in the band gap. As the impurity level shifts from above $E_\mathrm{F}$ before ionization [Fig.~\ref{fig:S2}(a)] to below $E_\mathrm{F}$ [Fig.~\ref{fig:S2}(b)], the impurity is ionized, further activating a local Colomb potential that lifts the bands. For bulk parabolic bands, this band lifting means a lower $\rho_\mathrm{s}$ in the conduction band at a fixed, positive bias, and thus a reduction in current \cite{Song2012-of}. However, our observation of a protrusion surrounding impurities in STM images (Fig. 1 and Fig.~\ref{fig:S1}) indicates a retraction of the tip in constant-current mode, which must be a response to the instantaneous increase in current. We believe that the $\rho_\mathrm{s}$ on the surface has a dominating contribution from surface resonance \cite{Han2014-bs, Golias2016-iq}, whose energy distribution has a peak close to the conduction band edge [Fig.~\ref{fig:S2}(c)]. When ionization takes place and the Coulomb potential lifts those bands [Fig.~\ref{fig:S2}(d)], this peak of $\rho_\mathrm{s}$ in the surface resonance, albeit not visible in the differential conductance spectrum in constant-height mode, aligns closer with the maximum of $T(E)$ (longer arrow). As a result, the current increase is possible when the tip moves from the position in Fig.~\ref{fig:S2}(a) to the position in Fig.~\ref{fig:S2}(b).

Second, as mentioned in the main text, the observed disk feature indicates a different ionization state from those outside it. More importantly, the edge of the disk outlines an equipotential contour for band bending. By solving Poisson's equation, one could reconstruct the three-dimensional electrostatic potential around the defect, and determine the dependence of the TIBB disk’s diameter as a function of bias voltage and/or tip-sample distance \cite{Feenstra1987-du}.

To capture the physics, we focus on a simplified model \cite{Battisti2017-hp} of a charged sphere above a semi-infinite dielectric material, the band bending $\varphi_\mathrm{BB}$ is determined by the sample bias voltage $V_\mathrm{B}$, the radius of the tip apex $r$, the dielectric constant of the sample (surface) $\varepsilon$, the tip-sample distance $d$, and the difference in work function of the tip $W_\mathrm{t}$ and sample $W_\mathrm{s}$,
\begin{equation}
\varphi_\mathrm{BB} = \frac{eV_\mathrm{B}+W_\mathrm{t}-W_\mathrm{s}}{1+\varepsilon d/r}.
\tag{S3}\label{eq:S3}
\end{equation}
Because the exact values of the above parameters cannot be precisely determined, in particular the tip’s work function $W_\mathrm{t}$ that may deviate from the value of a bulk material due to surface orientation, roughness, and absorption of alien atoms or molecules, the absolute value of $\varphi_\mathrm{BB}$ can vary within a range of 100 meV \cite{Teichmann2008-fn}. Such a large variation in $\varphi_\mathrm{BB}$ is comparable to or even larger than the typical impurity’s level spacing near $E_\mathrm{F}$ \cite{Wijnheijmer2009-fn}. Hence, we discuss the qualitative behavior below: we first describe the $V_\mathrm{B}$-dependence of the disk diameter consistent with our observation in Figs. 2(a)--(c), and then we discuss our observed $d$-dependence of the disk diameter.

%Figure S3
\begin{figure}[tb]
\includegraphics{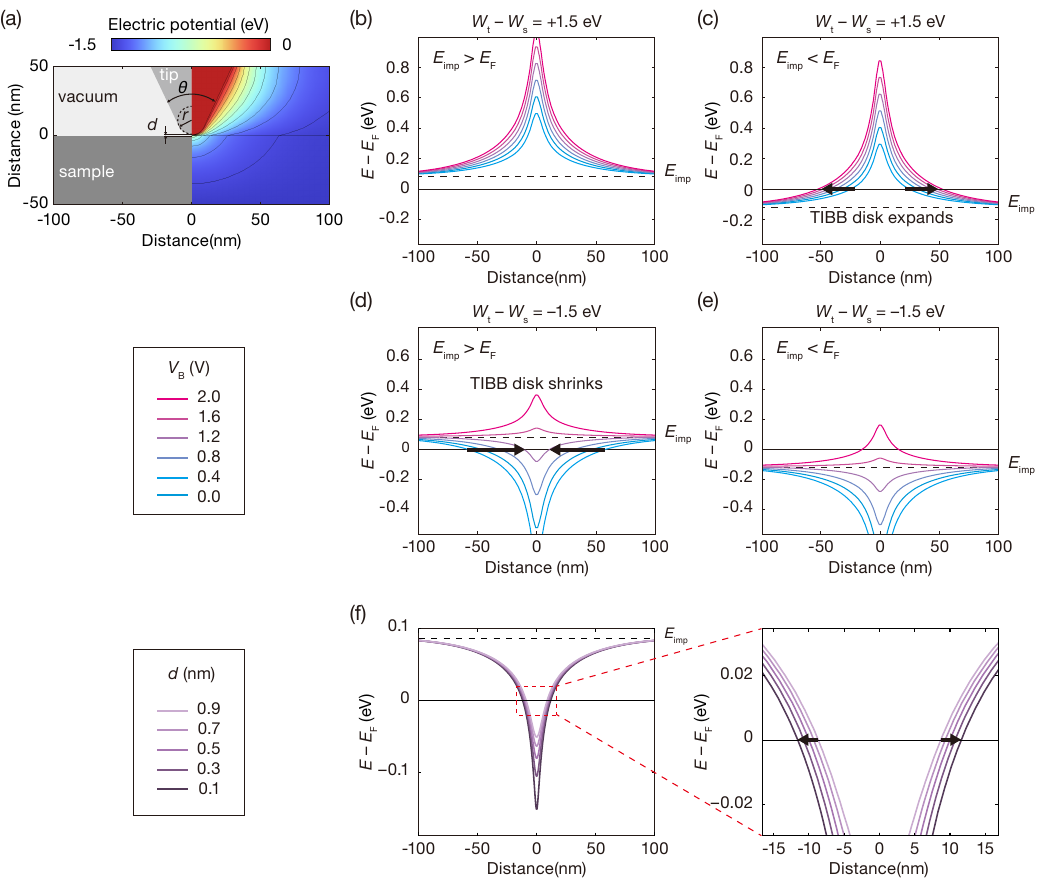}
\caption{Simulation of tip-induced band bending. (a) The model STM tunneling junction (left half) and the equipotential plot (right half) for a tip with an opening angle $\theta = 50^{\circ}$ and an apex radius $r$ = 10 nm, the tip-sample distance $d$ = 0.5 nm, the work function difference between the tip and the sample $W_\mathrm{t}– W_\mathrm{s}= –1.5$ eV, and a dielectric constant of the sample $\varepsilon$ = 13. (b)--(e) The calculated tip-induced band bending (TIBB) at the surface with increasing positive bias voltage at different conditions for the impurity level and $W_\mathrm{t}– W_\mathrm{s}$. (f) The calculated TIBB at the surface with decreasing tip-sample distance $d$ at a bias of 1.2 V (variation based on the purple curve in (d). The black dashed lines represent the impurity levels ($E_\mathrm{imp}$) without the tip effect. Colored curves correspond to $E_\mathrm{imp}$ in the presence of TIBB at different bias voltages (or tip-sample distances). For those curves crossing $E_\mathrm{F}$, their intersections at $E_\mathrm{F}$ mark the disk diameter within which the ionization occurs. The black arrows indicate changes in the diameter of the TIBB disk with increasing bias in (b)--(e) or decreasing tip-sample distance in (f).}
\label{fig:S3}
\end{figure}
Figure~\ref{fig:S3}(a) shows the cross-section plot of the electrical potential from a finite-element analysis simulation using COMSOL software. The tip is modeled by a cone ending with a hemisphere above the sample. The origin is set to the point on the surface right below the very end of the tip. We fix the opening angle of the cone to be 50$^{\circ}$, and exaggerate the difference between work functions $W_\mathrm{t}$ and $W_\mathrm{s}$ to be $\pm$ 1.5 eV for a clear qualitative illustration. At different positive bias $V_\mathrm{B} > 0$ we plot the band bending on the surface, with  the dashed line indicating the impurity level closest to the Fermi level $E_\mathrm{imp}$ without the tip effect. Based on signs of $E_\mathrm{imp} - E_\mathrm{F}$ and $W_\mathrm{t} - W_\mathrm{s}$, there are in total four different situations for band bending. For $W_\mathrm{t} > W_\mathrm{s}$, we always get $\varphi_\mathrm{BB} > 0$ from Eq.~\ref{eq:S1}. For an unoccupied impurity level $E_\mathrm{imp} > E_\mathrm{F}$, it keeps increasing and does not cross $E_\mathrm{F}$ [Fig.~\ref{fig:S3}(b)], therefore one does not expect to see the TIBB disk. For an occupied impurity level $E_\mathrm{imp} < E_\mathrm{F}$ [Fig.~\ref{fig:S3}(c)], it crosses $E_\mathrm{F}$  and the TIBB disk is bias-dependent: a larger bias results in a larger disk diameter. However, this dependence is opposite to what we observed in Figs. 2(a)--(c). When the sign of work function difference is reversed, i.e., $W_\mathrm{t} < W_\mathrm{s}$, Eq.~\ref{eq:S1} implies a 'flat-band' condition $V_\mathrm{B} = V_\mathrm{FB}$ that $\varphi_\mathrm{BB} = 0$. At this bias $V_\mathrm{FB}$ (between 1.2 V and 1.6 V in Figs.~\ref{fig:S3}(d) and \ref{fig:S3}(e), the impurity level remains constant such that the tip effect is absent. For an occupied impurity level, the TIBB disk only appears when $\lvert\varphi_\mathrm{BB}\rvert > \lvert E_\mathrm{imp}\rvert$ [Fig.~\ref{fig:S3}(e)], and the disk diameter increases with bias, same as Fig.~\ref{fig:S3}(c) and opposite to that in Figs. 2(a)--(c). For an unoccupied impurity level with $W_\mathrm{t} < W_\mathrm{s}$, one expects a TIBB disk shrinking in diameter with increasing bias for a negative $\varphi_\mathrm{BB}$ with $\lvert\varphi_\mathrm{BB}\rvert > \lvert E_\mathrm{imp}\rvert$ [Fig.~\ref{fig:S3}(d)]. Further increasing the bias leads to the disappearance of the TIBB disk and eventually reaching the flat-band condition. Only in this condition, the expectation is consistent with the experimental results in Figs. 2(a)--(c). The same dependence at positive bias was indeed observed in previous reports \cite{Song2012-of, Nguyen2018-tz}, while the opposite dependence is more commonly observed.

Next, we discuss the dependence of $\varphi_\mathrm{BB}$ on the tip-sample distance $d$. From Eq. S1 we find $\lvert\varphi_\mathrm{BB}\rvert$ increases with decreasing $d$, which can be interpreted by a higher electric field of the sample surface feels at a closer tip-sample distance. As a consequence, at a fixed bias where the TIBB disk is present, decreasing $d$ leads to a larger disk diameter \cite{Song2012-of}. Our simulation results in Fig.~\ref{fig:S2}(f) confirm this dependence, which, however, is opposite to what we observed in Figs. 2(d)--(f). As far as we know, this opposite dependence on tip-sample distance has not been reported before.

Nevertheless, from this $d$ dependence differential conductance maps (Figs.~\ref{fig:S5} and \ref{fig:S6}) we can confirm this effect originates from the tip’s electric field as the tip-sample distance (current setpoint) is the only varying parameter. It is also important to note that, for black phosphorus (BP), the TIBB effect on the apparent size of the band gap is minimal with changing current setpoint \cite{Kiraly2017-fg}, in contrast to a larger apparent gap with decreasing tip-sample distance as expected \cite{Battisti2017-hp}. Such an anomalous TIBB effect may result in the $V_\mathrm{B}$- and $d$-dependences less often seen. Whether the tip field has an extra effect on the local density of states other than band bending is yet to be investigated.

We finally emphasize that this anomalous effect is observed around all measured surface impurities (Fig.~\ref{fig:S1}). For cluster impurities, we arrive at the same observation. We note that the disk-like TIBB protrusions around nanoislands are not circular but anisotropic. For nanoislands, the situation becomes more complicated due to the Coloumb blockade effect, as indicated by differential conductance modulations on the islands \cite{Woodside2002-km}. The Coloumb blockade effect further depends on the size and shape of the nanoislands \cite{Brun2012-vu}, and the competition between the Coloumb blockade effect on the nanoisland and the band-bending effect around the nanoisland requires extensive theoretical modeling, which is beyond the scope of this study.

\section{Supplementary Note 2: Fourier filtering}
We applied Fourier filtering of STM images in Fig. 1(f) and Fig. 2 in the main text. Below we describe how exactly the Fourier filtering is carried out.
%Figure S4
\begin{figure}[tb]
\includegraphics{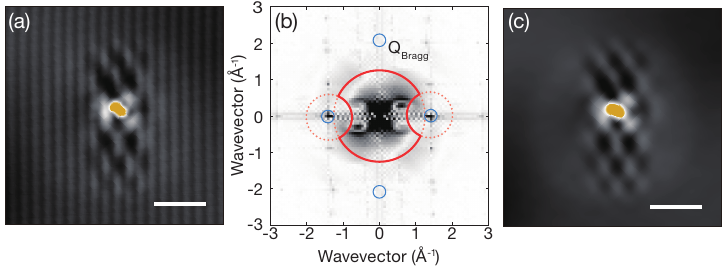}
\caption{Fourier filtering. (a) An STM image of the In clusters on the BP surface. Setup conditions: $V_\mathrm{B}$ = 0.8 V, $I_\mathrm{set}$ = 0.1 nA. (b) The amplitude of the Fourier transform of (a). The blue circles indicate the Bragg peaks of the BP surface. The solid closed curve in red mark the boundary within which the signal is passed by a smoothed step function (see Supplementary Note 2 for details). (c) The Fourier-filtered STM image of (a). Scale bar, 2 nm.}
\label{fig:S4}
\end{figure}

First, we note that the alternating bright and dark chain-like structure observed in Fig. 1(e) and Fig.~\ref{fig:S4}(a) comes from the crystal structure of the (001) surface of BP, and the zigzag atomic arrangement of the chains has been confirmed in previous reports \cite{Kiraly2017-fg, Kiraly2019-cg, Fang2022-my}, which requires an extremely sharp tip. Nevertheless, after a Fourier transform [the amplitude shown in Fig.~\ref{fig:S4}(b)], Bragg peaks can be resolved (blue circles), corresponding to lattice constants of 3.3 \AA~ and 4.4 \AA~ along the zigzag and armchair directions, respectively. Our results are in good agreement with previous reports.

\begin{figure}[h]
\includegraphics{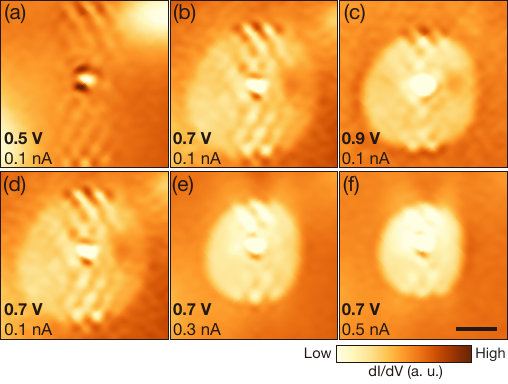}
\caption{Differential conductance maps with various tunneling conditions. The differential conductance maps are simultaneously acquired with the STM images shown in Fig. 2. To create the filtered images, we employed the procedures outlined in Supplementary Note 2. Scale bar, 2 nm.}
\label{fig:S5}
\end{figure}

\begin{figure}[htb]
\includegraphics{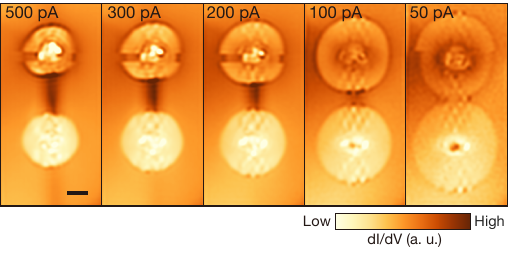}
\caption{The differential conductance maps are simultaneously acquired with the STM images shown in Fig. 4. To create the filtered images, we employed the procedures outlined in Supplementary Note 2. Scale bar, 2 nm.}
\label{fig:S6}

\end{figure}

Next, to better visualize the charge modulations, which are superimposed on the zigzag chains, we use a certain Fourier mask, as shown by the red curve in Fig.~\ref{fig:S4}(b). The mask excludes Bragg-peak signals of both zigzag and armchair directions and preserves all low-frequency (low-wavevector) signals. The periodic nature of the charge modulations is manifested by the oval-shaped peaks within the red curve. To create such a mask, one first filters out all the signal outside a radius of 1.3 \AA$^{-1}$ (2$\pi$/4.8 \AA$^{-1}$) from the origin, then further filters out signals within a radius of 0.6 \AA$^{-1}$ around two Bragg peaks (red dashed circles) along armchair directions. The circular masks are realized with a two-dimensional step function. The step function is smoothed by a Gaussian function with a full width at half maximum (FWHM) of 0.44 \AA$^{-1}$ to avoid artifacts during the inverse Fourier transform. This Fourier mask is applied to both real and imaginary parts of the Fourier transform of the original image. Then we perform an inverse Fourier transform to the masked image (of complex numbers) and take the real part of the resulting image [Fig.~\ref{fig:S4}(c)]. The imaginary part has a very small amplitude (on average $10^{-15}$ of the standard deviation of the real part), indicating a successful filtering after inverse Fourier transform, and thus can be neglected. The above procedures are applied to generate all the Fourier-filtered images in Fig. 2, Fig.~\ref{fig:S5}, and Fig.~\ref{fig:S6}.

\section{Supplementary Note 3: screened Coulomb potential}
At the interface of a dielectric and vacuum, two-dimensional electrons screen an external charge, e.g., on a surface impurity, resulting in a Coulomb-like potential \cite{Teichmann2008-fn, Ebert1996-zh}
\begin{equation}
\varphi (R) = \frac{Qe}{4\pi\varepsilon^* R}\exp({-R/R_\mathrm{B}}).
\tag{S4}\label{eq:S4}
\end{equation}
where $Q$ is the external charge, $R$ is the distance from the external charge, $R_\mathrm{B}$ is the bulk screening length, and $\varepsilon^*=(\varepsilon_\mathrm{BP}+\varepsilon_0)/2$ is the classic approximation of the dielectric constant at the surface 16, i.e., the average dielectric constant of vacuum $\varepsilon_0$ and BP $\varepsilon_\mathrm{BP} = 13\varepsilon_0$ \cite{Tian2019-er}. When extracting the potential, we determine $R$ by $R = \sqrt{d^2+x^2}$, where $d$ is the tip-sample distance and $x$ is the lateral distance [horizontal axis of Fig. 3(b)]. For the purple curve in Fig. 3(c), we assume $d$ = 0.5 nm, $Q=2e$,and $R_\mathrm{B}$ = 4.5 nm \cite{Low2014-py}.

%Figure S5
%Figure S6

\section{Supplementary Note 4: Details on QPI simulations for asymmetrical and non-local impurities}

\begin{figure}[htb]
\includegraphics{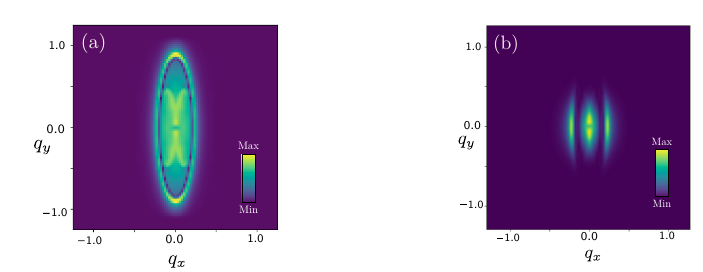}
\caption{QPI simulations for the  case (a) without form factors and local impurity, and (b) with form factors and non-local impurity. Both were obtained with $\omega=0.32E_{0}$ and $\mu=0.35E_{0}$. See text for more details.}
\label{fig:S7}

\end{figure}

In this section, we show how the presence of form-factors or quantum geometry can modify quasiparticle interference (QPI) patterns. To probe this, we calculate the local density of states in the presence of an impurity potential and then perform a Fourier transform (FT) to obtain the QPI. In a first-order approximation, the change in the local density of states for energy $\omega$ at position $\vec{r}$ is given by
\begin{equation}
    \delta\rho(\vec{r},\omega) = \frac{1}{\pi}\text{Im}\left[ \sum_{\vec{k}}\sum_{\vec{k}'}e^{i(\vec{k}-\vec{k}')\cdot\vec{r}}\text{Tr}\left[G_{0}(\vec{k},\omega)V_{\text{imp}}(\vec{k},\vec{k}')G_{0}(\vec{k}',\omega)\right]\right],
\end{equation}
where $G_{0}(\vec{k})$ is the bare electronic Green's function  and $V_{\text{imp}}(\vec{k},\vec{k}')$ the impurity potential which scatters an electron from $\vec{k}$ to $\vec{k}'$. Following \cite{Asahina1982Apr,zouFriedel} we can express the BP Hamiltonian (in the dimensionless form) as 
\begin{equation}
    H =\left(\frac{k_{x}^2}{2 m_{x}}+\frac{k_{y}^2}{2 m_{y}}\right)\tau_{z} + v k_x \tau_x-\mu \tau_{0},
    \label{eq:H}
\end{equation}
where $\tau_{0}$ is the identity matrix, and $\tau_{x,y,z}$ are the Pauli matrices in orbital space. We measure energy in the units of $E_{0} = \hbar^2/(2m_{e}a^2)$ with $a$ being the lattice parameter along $x$ direction and $m_{e}$ the electron mass. We also choose  $m_x = 0.07 m_y$ which leads to an anisotropic Fermi surface elongated along the $y$ direction, characteristic of BP \cite{Asahina1982Apr}. 
We first analyze the case where the potential is local in space and the quantum geometry of the wave functions are not taken into account. In this case, the dependence of the Bloch wave functions drops out, and the local density of states is simply given by
\begin{equation}
    \delta \rho(\vec{r},\omega) = \frac{1}{\pi}\text{Im}\left[\sum_{\vec{k},\vec{k}'} e^{i(\vec{k}-\vec{k}')\cdot \vec{r}}\left(\frac{1}{\omega + i \eta -\mathcal{E}_{\vec{k}}}\frac{1}{\omega + i \eta -\mathcal{E}_{\vec{k}'}}\right)\right],
    \label{eq:LDOS_withoutFF}
\end{equation}
 where $\mathcal{E}_{\vec{k}}$ are the energy eigenvalues of $H$ in \equref{eq:H} and $\eta = 10^{-2} E_{0}$ is a broadening parameter. Performing a Fourier transform leads to the QPI pattern shown in \figref{fig:S7}(a). Given \equref{eq:LDOS_withoutFF}, the dominant contribution to LDOS comes from the poles of the Green's function, and therefore most of the scatterings occur from regions on the Fermi surface.
 
To go beyond the isotropic case, we consider a non-local potential form with selective orbital coupling given by 
\begin{equation}
    V_{\text{imp}}(\vec{k},\vec{k}') = V_{0}(\tau_{0} -\tau_z),
\end{equation}
where $V_{0} = e^{-(\vec{k}-\vec{k}')^2/\sigma^{2}}$ encodes the non-point like nature of the impurity, facilitating scatterings from $\vec{k}$ to $\vec{k}'$. We choose $\sigma = 0.6$ in units of $1/a$. Additionally, we also include $\vec{k}$ dependent Bloch wave-functions $\phi(\vec{k})$ in our Green's function and calculate FT[$\rho(\vec{r,\omega})$]. We see (in \figref{fig:S7}(b)) that on account of the form factors, non-local nature of the impurity and asymmetrical coupling the QPI pattern undergoes severe reconstructions. In fact, we see that for this choice of $V_{\text{imp}}$, scatterings along $q_y$ are heavily suppressed. %In real space, this would translate to a distortion of the expected Friedel oscillations in real space for BP \cite{Kiraly2019-cg,zouFriedel}. 
%These results highlight how the QPI patterns are dependent on the nature of impurity (i.e., form of the impurity potential), selective couplings and quantum geometry of the material considered, besides the bare electronic dispersion.  

These results highlight how the complex interplay between non-local impurity potentials, selective orbital coupling, and quantum geometric factors introduces additional complexity to QPI patterns. This raises the possibility that such effects could play a role in the apparent discrepancies between experimental observations (see Fig. 4d of the main text) and expectations from the bare band structure.

\bibliography{ref}